\documentclass[twocolumn]{revtex4}

\usepackage{graphicx}% Include figure files
\usepackage{dcolumn}% Align table columns on decimal point
\usepackage{amstext,amsmath,amssymb,amsfonts,braket}
\usepackage{bm}% bold math

\usepackage{txfonts}
\usepackage{hyperref}
\usepackage{color}
\definecolor{lightblue}{rgb}{0.2,0.2,0.7}
\definecolor{darkblue}{rgb}{0,0.25,0.5}
\definecolor{redbrown}{rgb}{0.875,0.25,0.125}
\definecolor{darkgreen}{rgb}{0,0.5,0}

\renewcommand{\b}[1]{\ensuremath{\mathbf{#1}}}
\renewcommand{\H}{\ensuremath{\text{H}}}
\renewcommand{\l}{\ensuremath{\lambda}}

\newcommand{\tr}{\ensuremath{\text{tr}}}
\newcommand{\lr}{\ensuremath{\text{lr}}}
\newcommand{\sr}{\ensuremath{\text{sr}}}
\newcommand{\ee}{\ensuremath{\text{ee}}}

\newcommand{\HF}{\ensuremath{\text{HF}}}

\newcommand{\RPAx}{\ensuremath{\text{RPAx}}}

\newcommand{\RPAxSOt}{\ensuremath{\text{RPAxSO2}}}

\newcommand{\CCD}{\ensuremath{\text{CCD}}}

\renewcommand{\d}{\ensuremath{\text{d}}}

\newcommand{\x}{\ensuremath{\text{x}}}
\newcommand{\xc}{\ensuremath{\text{xc}}}
\renewcommand{\c}{\ensuremath{\text{c}}}
\newcommand{\Hxc}{\ensuremath{\text{Hxc}}}

\DeclareMathOperator{\erf}{erf}
\DeclareMathOperator{\erfc}{erfc}

\newcommand{\Cs}{\ensuremath{\text{C}_\text{s}}}
\newcommand{\Cdh}{\ensuremath{\text{C}_{2\text{h}}}}
\newcommand{\Cdv}{\ensuremath{\text{C}_{2\text{v}}}}
\newcommand{\Ctv}{\ensuremath{\text{C}_{3\text{v}}}}
\newcommand{\Dtd}{\ensuremath{\text{D}_{3\text{d}}}}
\newcommand{\Ddh}{\ensuremath{\text{D}_{2\text{h}}}}
\makeatletter
\renewcommand\paragraph{\@startsection{paragraph}{4}{\z@}%
  {-3.25ex\@plus -1ex \@minus -.2ex}%
  {1.5ex \@plus .2ex}%
  {\normalfont\normalsize\bfseries}}
\makeatother

\begin{document}

\title{Range-separated double-hybrid density-functional theory with coupled-cluster and random-phase approximations}

\author{Cairedine Kalai$^1$}
\author{Bastien Mussard$^2$}
\author{Julien Toulouse$^1$}\email{toulouse@lct.jussieu.fr}
\affiliation{$^1$Laboratoire de Chimie Th\'eorique (LCT), Sorbonne Universit\'e and CNRS, F-75005 Paris, France\\
$^2$Department of Chemistry and Biochemistry, University of Colorado Boulder, CO 80302 Boulder, USA}

\date{April 30, 2019}

\begin{abstract}
We construct range-separated double-hybrid schemes which combine coupled-cluster or random-phase approximations with a density functional based on a two-parameter Coulomb-attenuating-method-like decomposition of the electron-electron interaction. We find that the addition of a fraction of short-range electron-electron interaction in the wave-function part of the calculation is globally beneficial for the range-separated double-hybrid scheme involving a variant of the random-phase approximation with exchange terms. Even though the latter scheme is globally as accurate as the corresponding scheme employing only second-order M{\o}ller-Plesset perturbation theory for atomization energies, reaction barrier heights, and weak intermolecular interactions of small molecules, it is more accurate for the more complicated case of the benzene dimer in the stacked configuration. The present range-separated double-hybrid scheme employing a random-phase approximation thus represents a new member in the family of double hybrids with minimal empiricism which could be useful for general chemical applications.
\end{abstract}

\maketitle

\section{Introduction}

In density-functional theory (DFT) of molecular electronic systems, the last decade has seen the emergence of double-hybrid approximations~\cite{Gri-JCP-06} (see Refs.~\onlinecite{SanAda-PCCP-13,GoeGri-WIRE-14,BreCioSanAda-ACR-16,MehCasGor-PCCP-18} for reviews). These approaches combine Hartree-Fock (HF) exchange and second-order M{\o}ller--Plesset (MP2) correlation with a semilocal exchange-correlation density-functional approximation (DFA) based on a linear separation of the Coulomb electron-electron interaction~\cite{ShaTouSav-JCP-11}. These double-hybrid approximations have the advantage of having quite small self-interaction error~\cite{SuYanMorXu-JPCA-14} thanks to their large fraction of HF exchange. Alternatively, range-separated density-functional theory~\cite{Sav-INC-96,TouColSav-PRA-04} also provides a way for combining a correlated wave-function method with a semilocal DFA based on a separation of the electron-electron interaction into long-range and short-range contributions. For example, long-range HF exchange and long-range MP2 correlation can be combined with a short-range semilocal exchange-correlation DFA~\cite{AngGerSavTou-PRA-05}, with the advantage of explicitly describing long-range van der Waals dispersion interactions~\cite{TayAngGalZhaGygHirSonRahLilPodBulHenScuTouPevTruSza-JCP-16}.

Recently, a range-separated double-hybrid (RSDH) approximation~\cite{KalTou-JCP-18} has been constructed based on the following decomposition of the Coulomb electron-electron interaction $w_\ee(r_{12})=1/r_{12}$
\begin{eqnarray}
w_\ee(r_{12}) &=& \left[ w_\ee^{\lr,\mu} (r_{12}) + \l w_\ee^{\sr,\mu} (r_{12}) \right]
\nonumber\\
&&+ (1-\l) w_\ee^{\sr,\mu} (r_{12}),
\label{decompRSDH}
\end{eqnarray}
where $w_\ee^{\lr,\mu} (r_{12}) = \erf(\mu r_{12})/r_{12}$ is a long-range interaction (written with the error function erf), $w_\ee^{\sr,\mu} (r_{12}) = \erfc(\mu r_{12})/r_{12}$ is the complementary short-range interaction (written with the complementary error function erfc), and $\mu$ and $\l$ are two parameters. The first term in the square bracket in Eq.~(\ref{decompRSDH}) is treated by MP2 and the remaining term by a semilocal DFA. This RSDH approximation generalizes the double hybrids (corresponding to the special case $\mu=0$) and the range-separated hybrids (corresponding to the special case $\l=0$). The advantage of the RSDH approximation is that long-range interactions are explicitly described while the addition of a fraction of short-range interaction in the wave-function part of the calculation reduces the self-interaction error. The two-parameter decomposition of Eq.~(\ref{decompRSDH}) has also been used to combine pair coupled-cluster doubles with a semilocal DFA for describing static correlation~\cite{GarBulHenScu-PCCP-15}. This decomposition is in fact a special case of the three-parameter decomposition used in the Coulomb-attenuating method (CAM)~\cite{YanTewHan-CPL-04} which has also been considered for constructing double-hybrid approximations~\cite{CorFro-IJQC-14}. Other related double-hybrid approximations have been proposed which combine a long-range HF exchange term with a full-range MP2 correlation term~\cite{ChaHea-JCP-09}, a full-range HF exchange term with a long-range MP2 correlation term~\cite{BenDisLocChaHea-JPCA-08,ZhaXu-JPCL-13}, or a HF exchange term based on the decomposition of Eq.~(\ref{decompRSDH}) with a full-range MP2 correlation term~\cite{BreSavPerSanAda-JCTC-18}.

In this work, we extend the RSDH approximation based on the decomposition of Eq.~(\ref{decompRSDH}) by supplanting the MP2 correlation term by a random-phase-approximation (RPA) correlation term. Specifically, we use the so-called RPAxSO2 variant~\cite{SzaOst-JCP-77,TouZhuSavJanAng-JCP-11,MusReiAngTou-JCP-15} which is a RPA with exchange terms. In the context of the range-separated hybrids, it was shown that long-range RPAxSO2 gives intermolecular interaction energies overall more accurate than long-range MP2, especially for dispersion-dominated large molecular complexes~\cite{TouZhuSavJanAng-JCP-11,TayAngGalZhaGygHirSonRahLilPodBulHenScuTouPevTruSza-JCP-16}. We thus expect a similar improvement in the context of the RSDH approach. Since RPAxSO2 is a simplification of coupled cluster doubles (CCD), we also test the use of CCD in the RSDH approach. Let us mention that a number of double hybrids using the direct RPA approximation (without exchange terms)~\cite{RuzPerCso-JCTC-10,AhnHehVogTraLeuKlo-CP-14,MezCsoRuzKal-JCTC-15,GriSte-PCCP-16,MezCsoRuzKal-JCTC-17} or coupled-cluster approximations~\cite{ChaGoeRad-JCC-16} have already been proposed, as well as range-separated hybrids using various RPA variants~\cite{TouGerJanSavAng-PRL-09,JanHenScu-JCP-09,JanHenScu-JCP-09b,ZhuTouSavAng-JCP-10,TouZhuAngSav-PRA-10,PaiJanHenScuGruKre-JCP-10,TouZhuSavJanAng-JCP-11,AngLiuTouJan-JCTC-11,IreHenScu-JCP-11,CheMusAngRei-CPL-12,MusSzaAng-JCTC-14,MusReiAngTou-JCP-15,GarBulSouSunPerScu-MP-16,HesAng-TCA-18} or coupled-cluster approximations~\cite{GolWerSto-PCCP-05,GolWerStoLeiGorSav-CP-06,GolStoThiSch-PRA-07,GolWerSto-CP-08,GolErnMoeSto-JCP-09,TouZhuSavJanAng-JCP-11}. However, to the best of our knowledge, the use of RPA or coupled-cluster approximations with the general decomposition of Eq.~(\ref{decompRSDH}) had never been tried before.

The paper is organized as follows. In Section~\ref{sec:theory}, the theory underlying the RSDH scheme with coupled-cluster and RPA approximations is presented. Computational details are given in Section~\ref{sec:comp}. In Section~\ref{sec:results}, we give and discuss the results, including the optimization of the parameters $\mu$ and $\l$ on small sets of atomization energies (AE6 set) and reaction barrier heights (BH6 set), tests on larger sets of atomization energies (AE49 set), reaction barrier heights (DBH24 set), weak intermolecular interactions (A24 set), and on the interaction energy curve of the benzene dimer in the stacked configuration. Section~\ref{sec:conclusion} contains conclusions. Unless otherwise specified, Hartree atomic units are tacitly assumed throughout this work.

\section{THEORY}
\label{sec:theory}

In the RSDH approach, the exact ground-state electronic energy of a $N$-electron system is expressed as~\cite{KalTou-JCP-18}
\begin{eqnarray}
E &=& \min_{\Psi} \Bigl\{ \bra{\Psi} \hat{T} + \hat{V}_\text{ne} + \hat{W}_\ee^{\lr,\mu} + \l \hat{W}_\ee^{\sr,\mu}  \ket{\Psi}
\nonumber\\
&& \;\;\;\;\;\;\;\;\;\; + \bar{E}_\Hxc^{\sr,\mu,\l}[n_{\Psi}] \Bigl\},
\label{GS-energy1}
\end{eqnarray}
where the minimization is done over $N$-electron normalized multideterminant wave functions $\Psi$. In Eq.~(\ref{GS-energy1}), $\hat{T}$ is the kinetic-energy operator, $\hat{V}_\text{ne}$ is the nuclei-electron potential operator, $\hat{W}_\ee^{\lr,\mu} = (1/2) \iint w_\ee^{\lr,\mu} (r_{12}) \hat{n}_{2}(\textbf{r}_{1},\textbf{r}_{2}) \d\b{r}_{1} \d\b{r}_{2}$ and $\hat{W}_\ee^{\sr,\mu} = (1/2) \iint w_\ee^{\sr,\mu} (r_{12}) \hat{n}_{2}(\textbf{r}_{1},\textbf{r}_{2}) \d\b{r}_{1} \d\b{r}_{2}$ are the long-range and short-range electron-electron interaction operators (expressed with the pair-density operator $\hat{n}_{2}(\textbf{r}_{1},\textbf{r}_{2})$), respectively, and $\bar{E}_\Hxc^{\sr,\mu,\l}[n_{\Psi}]$ is the complement short-range Hartree-exchange-correlation density functional evaluated at the density of $\Psi$, i.e. $n_{\Psi}(\b{r})=\bra{\Psi} \hat{n}(\b{r})\ket{\Psi}$ where $\hat{n}(\b{r})$ is the density operator. The minimizing normalized wave function in Eq.~(\ref{GS-energy1}) will be denoted by $\Psi^{\mu,\l}$. It satisfies the nonlinear Schr\"odinger-like eigenvalue equation
\begin{eqnarray}
\hat{H}^{\mu,\l}[n_{\Psi^{\mu,\l}}] \ket{\Psi^{\mu,\l}} = {\cal E}^{\mu,\l} \ket{\Psi^{\mu,\l}},
\label{EL1}
\end{eqnarray}
with the Hamiltonian $\hat{H}^{\mu,\l}[n] = \hat{T} + \hat{V}_\text{ne} + \hat{W}_\ee^{\lr,\mu} + \l \hat{W}_\ee^{\sr,\mu} + \hat{V}_\Hxc^{\sr,\mu,\l}[n]$ which includes the complement short-range Hartree-exchange-correlation potential operator $\hat{V}_\Hxc^{\sr,\mu,\l}[n] = \int v_\Hxc^{\sr,\mu,\l}(\b{r}) \hat{n}(\b{r}) \d \b{r}$ with $v_\Hxc^{\sr,\mu,\l}(\b{r}) = \delta \bar{E}_\Hxc^{\sr,\mu,\l}[n]/\delta n(\b{r})$. It is assumed that the minimizing wave function $\Psi^{\mu,\l}$ in Eq.~(\ref{GS-energy1}) corresponds to the ground state of the self-consistent Hamiltonian $\hat{H}^{\mu,\l}[n_{\Psi^{\mu,\l}}]$. Then, by construction, the potential $v_\Hxc^{\sr,\mu,\l}(\b{r})$ ensures that the ground-state wave function $\Psi^{\mu,\l}$ gives the exact density, $n_{\Psi^{\mu,\l}}=n$, for all values of $\mu$ and $\l$.

As a first step, we use a single-determinant approximation in Eq.~(\ref{GS-energy1}), giving what we will call the range-separated two-parameter hybrid (RS2H) scheme,
\begin{eqnarray}
E^{\mu,\l}_\text{RS2H} &=& \min_{\Phi} \Bigl\{ \bra{\Phi} \hat{T} + \hat{V}_\text{ne} + \hat{W}_\ee^{\lr,\mu} + \l \hat{W}_\ee^{\sr,\mu}  \ket{\Phi} 
\nonumber\\
&& + \bar{E}_\Hxc^{\sr,\mu,\l}[n_{\Phi}] \Bigl\},
\label{RSH-1}
\end{eqnarray}
where the search is over $N$-electron normalized single-determinant wave functions $\Phi$. We will denote the minimizing RS2H normalized single-determinant wave function by $\Phi^{\mu,\l}$, whose density $n_{\Phi^{\mu,\l}}$, contrary to the one of $\Psi^{\mu,\l}$, is not the exact density. The exact ground-state energy can then be written as
\begin{eqnarray}
E = E^{\mu,\l}_\text{RS2H} + E^{\mu,\l}_\c,
\end{eqnarray}
where $E^{\mu,\l}_\c$ is the correlation energy associated with the interaction $w_\ee^{\lr,\mu} (r_{12}) + \l w_\ee^{\sr,\mu} (r_{12})$. Extending the work of Ref.~\onlinecite{TouZhuSavJanAng-JCP-11}, this correlation energy can be expressed as
\begin{eqnarray}
E^{\mu,\l}_\c &=& \bra{\Psi^{\mu,\l}} \hat{H}^{\mu,\l}[n] \ket{\Psi^{\mu,\l}} - \bra{\Phi^{\mu,\l}} \hat{H}^{\mu,\l}[n] \ket{\Phi^{\mu,\l}}
\nonumber\\
\label{Ecmul}
&& + \Delta \bar{E}_\Hxc^{\sr,\mu,\l} - \int \! v_\Hxc^{\sr,\mu,\l}[n](\b{r}) \; \Delta n(\b{r}) \d\b{r},
\end{eqnarray}
where the last two terms are the variation of the energy functional, $\Delta \bar{E}_\Hxc^{\sr,\mu,\l} = \bar{E}_\Hxc^{\sr,\mu,\l}[n] - \bar{E}_\Hxc^{\sr,\mu,\l}[n_{\Phi^{\mu,\l}}]$, and the variation of the associated potential expectation value due to the variation of the density from the RS2H one to the exact one, $\Delta n = n - n_{\Phi^{\mu,\l}}$. Alternatively, the correlation energy can be expressed with the projection formula
\begin{eqnarray}
E^{\mu,\l}_\c &=& \bra{\Phi^{\mu,\l}} \hat{H}^{\mu,\l}[n] \ket{\tilde{\Psi}^{\mu,\l}} - \bra{\Phi^{\mu,\l}} \hat{H}^{\mu,\l}[n] \ket{\Phi^{\mu,\l}}
\nonumber\\
\label{Ecmulproj}
&& + \Delta \bar{E}_\Hxc^{\sr,\mu,\l} - \int \! v_\Hxc^{\sr,\mu,\l}[n](\b{r}) \; \Delta n(\b{r}) \d\b{r},
\end{eqnarray}
using the intermediate-normalized wave function ${\tilde{\Psi}^{\mu,\l}} = {\Psi^{\mu,\l}} / \braket{\Phi^{\mu,\l}|\Psi^{\mu,\l}}$.

Up to here the theory was exact. Let us now introduce the CCD ansatz for the wave function
\begin{eqnarray}
\ket{\tilde{\Psi}^{\mu,\l}_{\CCD}} = \exp \left(\hat{T}_2\right) \ket{\Phi^{\mu,\l}},
\end{eqnarray}
where $\hat{T}_2 = (1/4) \sum_{ijab} t_{ij}^{ab} \hat{a}_a^\dag \hat{a}_i \hat{a}_b^\dag \hat{a}_j$ is the double-excitation cluster operator written in terms of the amplitudes $t_{ij}^{ab}$, and occupied ($i$, $j$) and virtual ($a$, $b$) RS2H spin-orbital creation and annihilation operators. We determine $t_{ij}^{ab}$ with the CCD amplitude equations
\begin{eqnarray}
\bra{\Phi^{\mu,\l}_{ij\to ab}} \hat{H}^{\mu,\l}[n_{\Phi^{\mu,\l}}] \ket{\tilde{\Psi}^{\mu,\l}_\text{CCD}} = 0,
\label{CCDeq}
\end{eqnarray}
where $\Phi^{\mu,\l}_{ij\to ab}$ are doubly excited determinants, and we have used the approximation of keeping the density fixed at the RS2H density $n_{\Phi^{\mu,\l}}$ in the Hamiltonian. Equation~(\ref{CCDeq}) leads to the usual quadratic CCD amplitude equations, replacing the normal Hamiltonian by the modified $\hat{H}^{\mu,\l}[n_{\Phi^{\mu,\l}}]$, which just corresponds to using the RS2H orbital energies and the two-electron integrals associated with the interaction $w_\ee^{\lr,\mu} (r_{12}) + \l w_\ee^{\sr,\mu} (r_{12})$ in the usual CCD amplitude equations. The correlation energy $E^{\mu,\l}_\c$ is then approximated as
\begin{eqnarray}
E^{\mu,\l}_{\c,\text{CCD}} &=& \bra{\Phi^{\mu,\l}} \hat{H}^{\mu,\l}[n_{\Phi^{\mu,\l}}] \ket{\tilde{\Psi}^{\mu,\l}_\text{CCD}} 
\nonumber\\
&&\;\;\;\;\;\;\;- \bra{\Phi^{\mu,\l}} \hat{H}^{\mu,\l}[n_{\Phi^{\mu,\l}}] \ket{\Phi^{\mu,\l}},
\label{EcCCD}
\end{eqnarray}
where again, in comparison with Eq.~(\ref{Ecmulproj}), the variation of the density has been neglected, i.e. $n \approx n_{\Phi^{\mu,\l}}$. This is a reasonable approximation since the quantity $\Delta \bar{E}_\Hxc^{\sr,\mu,\l} - \int \! v_\Hxc^{\sr,\mu,\l}[n](\b{r}) \; \Delta n(\b{r}) \d\b{r}$ in Eq.~(\ref{Ecmulproj}) is of second order in $\Delta n$~\cite{TouZhuSavJanAng-JCP-11}. In matrix notation, the CCD correlation energy can be calculated as
\begin{eqnarray}
E_{\c,\CCD}^{\mu,\l} = \frac{1}{2} \tr \left[ \b{K} \, \b{T} \right],
\label{EcCCDtr}
\end{eqnarray}
where $K_{ia,jb}=\langle ij |\hat{w}_\ee^{\lr,\mu} + \l \hat{w}_\ee^{\sr,\mu} | ab \rangle$ are matrix elements made of two-electron integrals associated with the interaction $w_\ee^{\lr,\mu} (r_{12}) + \l w_\ee^{\sr,\mu} (r_{12})$ and $T_{ia,jb}=t_{ij}^{ab}$ are the amplitude matrix elements. 

We also consider the ring-diagram approximation with exchange terms (or linear-response time-dependent HF). In this approximation, referred to as RPAx, the CCD amplitude equations simplify to the following Riccati matrix equation giving the RPAx amplitudes $\b{T}_{\RPAx}$~\cite{ScuHenSor-JCP-08}
\begin{equation}
\b{B}^* + \b{A}^{\!*}  \, \b{T}_{\RPAx} + \b{T}_{\RPAx} \, \b{A} + \b{T}_{\RPAx} \, \b{B} \, \b{T}_{\RPAx} = \b{0},
\label{TRPAx}
\end{equation}
where the matrices $\b{A}$ and $\b{B}$ are
\begin{eqnarray}
A_{ia,jb} &=& (\varepsilon_a-\varepsilon_i) \delta_{ij} \delta_{ab} 
\nonumber\\
 &&+\langle ib | \hat{w}_\ee^{\lr,\mu} + \l\hat{w}_\ee^{\sr,\mu} |aj \rangle - \langle ib | \hat{w}_\ee^{\lr,\mu} + \l\hat{w}_\ee^{\sr,\mu} |ja \rangle,
\nonumber\\
\label{}
\end{eqnarray}
and
\begin{equation}
B_{ia,jb}=\langle ij |\hat{w}_\ee^{\lr,\mu} + \l\hat{w}_\ee^{\sr,\mu} |ab \rangle - \langle ij |\hat{w}_\ee^{\lr,\mu} + \l\hat{w}_\ee^{\sr,\mu} |ba \rangle,
\label{}
\end{equation}
written in terms of the RS2H orbital energies $\varepsilon_k$ and the same two-electron integrals introduced above. Once the RPAx amplitudes are obtained, the RPAxSO2 correlation energy is calculated by~\cite{SzaOst-JCP-77,TouZhuSavJanAng-JCP-11,MusReiAngTou-JCP-15}
\begin{equation}
E_{\c,\RPAxSOt}^{\mu,\l} = \frac{1}{2} \tr \left[ \b{K} \, \b{T}_{\RPAx} \right].
\label{EcRPAxSOt}
\end{equation}
For closed-shell systems, we use a spin-restricted formalism and the RPAxSO2 method involves only spin-singlet excitations. For open-shell systems, we use a spin-unrestricted formalism and the RPAxSO2 method involves only non-spin-flipped excitations~\cite{MusReiAngTou-JCP-15}. 

It remains to specify the approximation used for the complement short-range Hartree-exchange-correlation density functional. We first decompose it as
\begin{equation}
\bar{E}_\Hxc^{\sr,\mu,\l}[n] = E_\H^{\sr,\mu,\l}[n] + E_\x^{\sr,\mu,\l}[n] +  \bar{E}_{\c}^{\sr,\mu,\l}[n],
\end{equation}
where the short-range Hartree and exchange contributions are linear in $\l$
\begin{eqnarray}
E_\H^{\sr,\mu,\l}[n] = (1-\l) E_\H^{\sr,\mu}[n],
\label{EHsrmul}
\end{eqnarray}
\begin{eqnarray}
E_\x^{\sr,\mu,\l}[n] &=& (1-\l) E_\x^{\sr,\mu}[n],
\label{Exsrmul}
\end{eqnarray}
where $E_\H^{\sr,\mu}[n]$ and $E_\x^{\sr,\mu}[n]$ are the Hartree and exchange functionals defined with the short-range interaction $w_\ee^{\sr,\mu} (r_{12})$~\cite{TouColSav-PRA-04,TouSav-JMS-06}. The $\l$ dependence in the complement short-range correlation functional is approximated as (referred to as ``approximation 3'' in Ref.~\onlinecite{KalTou-JCP-18})
\begin{eqnarray}
\bar{E}_{\c}^{\sr,\mu,\l}[n] \approx \bar{E}_\c^{\sr,\mu}[n] - \l^2 \bar{E}_\c^{\sr,\mu\sqrt{\l}}[n],
\label{Ecsrmul}
\end{eqnarray}
where $\bar{E}_\c^{\sr,\mu}[n]$ is the usual complement short-range correlation functional~\cite{TouColSav-PRA-04,TouSav-JMS-06}. The $\l$ dependence in Eq.~(\ref{Ecsrmul}) is correct both in the high-density limit, for a non-degenerate KS system, and in the low-density limit. Finally, for $E_\x^{\sr,\mu}[n]$ and $\bar{E}_\c^{\sr,\mu}[n]$, we use the short-range Perdew-Becke-Ernzerhof (PBE) exchange and correlation functionals of Ref.~\onlinecite{GolWerStoLeiGorSav-CP-06}.

To summarize, the exchange-correlation energy in what we will call the RS2H+CCD and RS2H+RPAxSO2 methods is 
\begin{eqnarray}
E^{\mu,\l}_{\xc,\text{RS2H+CCD/RPAxSO2}} = E_{\x,\HF}^{\lr,\mu} + \l E_{\x,\HF}^{\sr,\mu} \;\;\; \;\;\; \;\;\; \;\;\; \;\;\;
\nonumber\\
+ (1-\l) E_\x^{\sr,\mu}[n] + \bar{E}_{\c}^{\sr,\mu,\l}[n] + E^{\mu,\l}_{\c,\text{CCD/RPAxSO2}}. \;\;\;
\end{eqnarray}
Note that, at second-order in the electron-electron interaction, both CCD and RPAxSO2 correlation energy expressions reduce to the MP2 correlation energy expression, and thus the RS2H+CCD and RS2H+RPAxSO2 methods reduce to a method that we will refer to as RS2H+MP2. This RS2H+MP2 method exactly corresponds to the method that was referred to as ``RSDH with approximation 3'' in Ref.~\onlinecite{KalTou-JCP-18}. For $\l=0$, the RS2H+CCD, RS2H+RPAxSO2, and RS2H+MP2 methods reduce to the RSH+CCD~\cite{TouZhuSavJanAng-JCP-11}, RSH+RPAxSO2~\cite{TouZhuSavJanAng-JCP-11}, and RSH+MP2~\cite{AngGerSavTou-PRA-05} methods, respectively, while for $\l=1$ they reduce to full-range CCD, RPAxSO2, and MP2 (all with HF orbitals), respectively.

%%%%%%%%%%%%%%%%%%%%%%%%%%%%%%%%%%%%%%%%%%%%%%%%%%%%%%%%%%%%%%%%%%%%%%%%%%%%%%%%%%%%%%%%%%%%%%%%%%
\begin{figure*}[t]
\centering
\includegraphics[scale=0.50,angle=0]{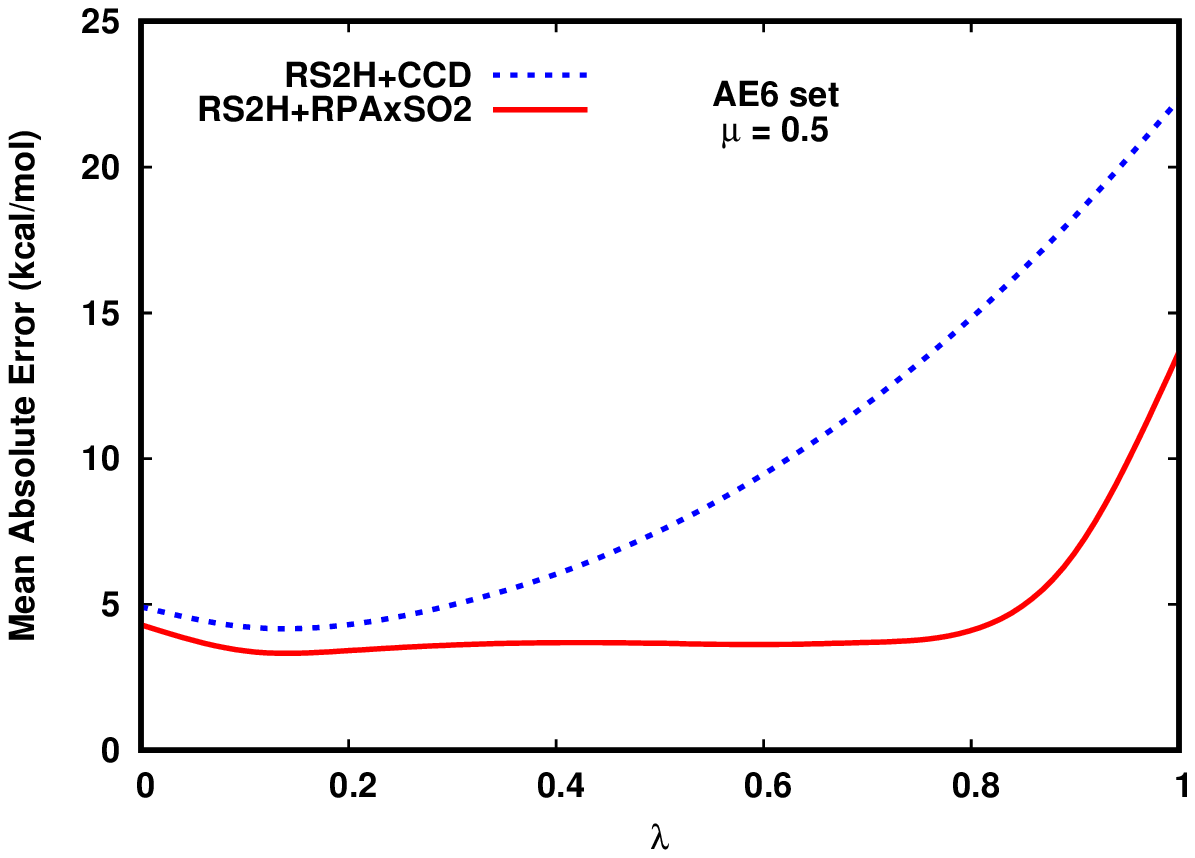}
\includegraphics[scale=0.50,angle=0]{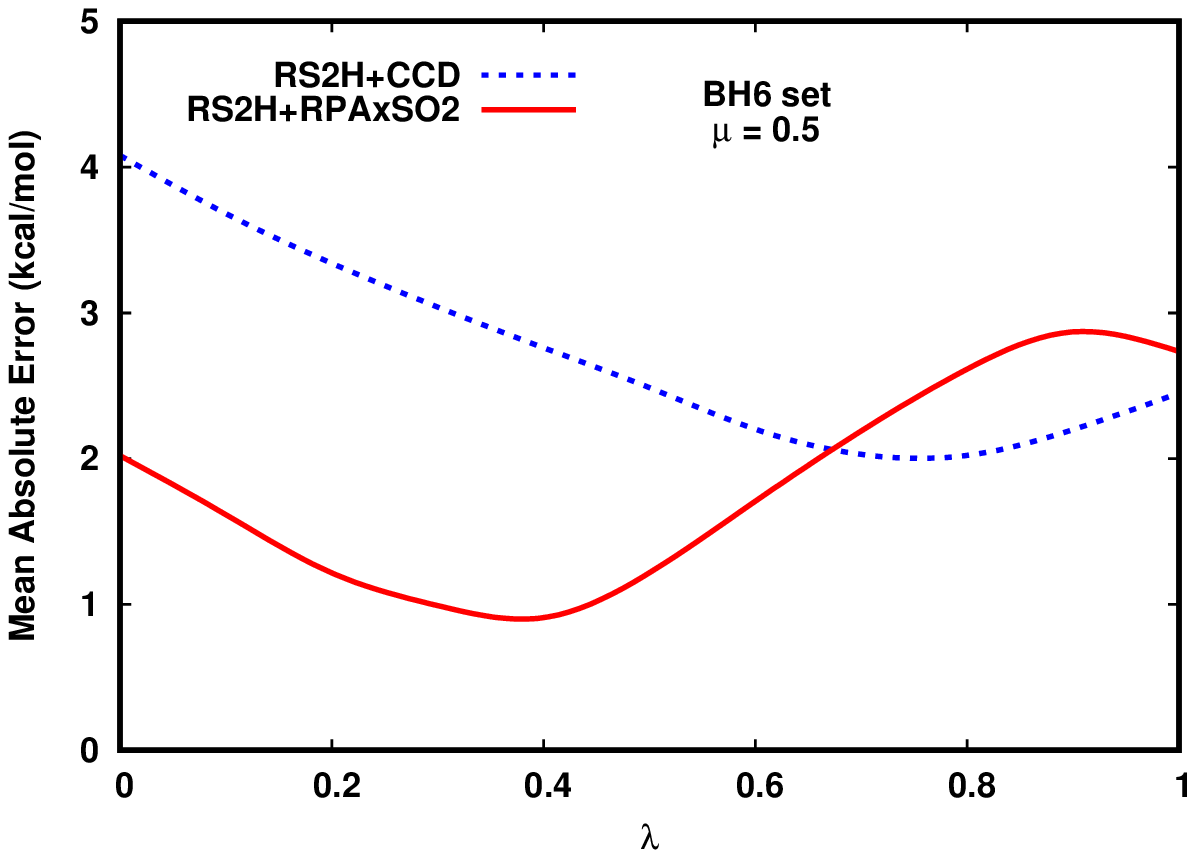}
\caption{MAEs for the AE6 and BH6 sets calculated with the RS2H+CCD and RS2H+RPAxSO2 methods as a function of $\l$ for $\mu=0.5$. The basis set used is cc-pVTZ.}
\label{AE6_BH6}
\end{figure*}
%%%%%%%%%%%%%%%%%%%%%%%%%%%%%%%%%%%%%%%%%%%%%%%%%%%%%%%%%%%%%%%%%%%%%%%%%%%%%%%%%%%%%%%%%%%%%%%%%%

\section{Computational details}
\label{sec:comp}

The RS2H+CCD and RS2H+RPAxSO2 methods have been implemented in a development version of the {\sc MOLPRO} program~\cite{Molproshort-PROG-19}. The calculation is done in two steps: first a self-consistent-field calculation is performed according to Eq.~(\ref{RSH-1}) and Eqs.~(\ref{EHsrmul})-(\ref{Ecsrmul}), and then the CCD or RPAxSO2 correlation energy in Eq.~(\ref{EcCCDtr}) or~(\ref{EcRPAxSOt}) is calculated using the previously obtained orbitals. 

The RS2H+CCD and RS2H+RPAxSO2 methods were applied on the AE6 and BH6 sets~\cite{LynTru-JPCA-03}, as a first assessment of the approximations on molecular systems and in order to determine the optimal parameters $\mu$ and $\l$. The AE6 set is a small representative benchmark of six atomization energies consisting of SiH$_4$, S$_2$, SiO, C$_3$H$_4$ (propyne), C$_2$H$_2$O$_2$ (glyoxal), and C$_4$H$_8$ (cyclobutane). The BH6 set is a small representative benchmark of forward and reverse hydrogen transfer barrier heights of three reactions, OH + CH$_4$ $\rightarrow$ CH$_3$ + H$_2$O, H + OH $\rightarrow$ O + H$_2$, and H + H$_2$S $\rightarrow$ HS + H$_2$. All the calculations for the AE6 and BH6 sets were performed with the Dunning cc-pVTZ basis set~\cite{Dun-JCP-89} at the geometries optimized by quadratic configuration interaction singles doubles with the modified Gaussian-3 basis set (QCISD/MG3)~\cite{KalTou-JJJ-XX-note1}. The reference values for the atomization energies and barrier heights are the non-relativistic frozen-core (FC) explicitly-correlated coupled-cluster singles doubles and perturbative triples [CCSD(T)]/cc-pVQZ-F12 values of Refs.~\onlinecite{HauKlo-TCA-12,HauKlo-TCA-12-err}. 

The RS2H+CCD and RS2H+RPAxSO2 methods were then tested on the AE49 set of 49 atomization energies~\cite{FasCorSanTru-JPCA-99} (consisting of the G2-1 set~\cite{CurRagTruPop-JCP-91,CurRagRedPop-JCP-97} stripped of the six molecules containing Li, Be, and Na) and on the DBH24/08 set~\cite{ZheZhaTru-JCTC-07,ZheZhaTru-JCTC-09} of 24 forward and reverse reaction barrier heights. These calculations were performed with the aug-cc-pVTZ basis set~\cite{WooDun-JCP-93}, with MP2(full)/6-31G* geometries for the AE49 set and QCISD/MG3 geometries for the DBH24/08 set. The reference values for the AE49 set are the non-relativistic FC CCSD(T)/cc-pVQZ-F12 values of Ref.~\onlinecite{HauKlo-JCP-12}, and the reference values for the DBH24/08 set are the zero-point exclusive values from Ref.~\onlinecite{ZheZhaTru-JCTC-09}.

The RS2H+CCD and RS2H+RPAxSO2 methods, as well as the RS2H+MP2 method, were also tested on the A24 set of 24 weakly interacting molecular complexes~\cite{RezHob-JCTC-13}. These calculations were performed with the aug-cc-pVTZ basis set and the counterpoise correction, using the composite complete-basis-set (CBS) CCSD(T) geometries of Ref.~\onlinecite{RezHob-JCTC-13} and the non-relativistic reference interaction energies from Ref.~\onlinecite{RezDubJurHob-PCCP-15}. Finally, the RS2H+CCD, RS2H+RPAxSO2, and RS2H+MP2 methods were compared on the interaction energy curve of the benzene dimer in the stacked (parallel-displaced) configuration. These calculations were performed with the aug-cc-pVDZ basis set and the counterpoise correction, using the geometries and the CCSD(T)/CBS reference interaction energies from the S66$\times$8 data set (item number 24)~\cite{RezRilHob-JCTC-11}.

Core electrons are kept frozen in all our CCD, RPAxSO2, and MP2 calculations. Spin-restricted calculations are performed for all the closed-shell systems, and spin-unrestricted calculations for all the open-shell systems. 

As statistical measures of accuracy of the different methods, we compute mean absolute errors (MAEs), mean errors (MEs), root mean square deviations (RMSDs), mean absolute percentage errors (MA\%E), and maximal and minimal errors.

%%%%%%%%%%%%%%%%%%%%%%%%%%%%%%%%%%%%%%%%%%%%%%%%%%%%%%%%%%%%%%%%%%%%%%%%%%%%%%%%%%%%%%%%%%%%%%%%%%%%%%%%%%%%%%%%
\begingroup
\squeezetable
\begin{table*}
\renewcommand{\arraystretch}{0.8}
\setlength{\tabcolsep}{0.07cm}
\footnotesize
\caption{Atomization energies (in kcal/mol) of the AE49 set calculated by RSH+CCD, RS2H+CCD, CCD, RSH+RPAxSO2, RS2H+RPAxSO2, and RPAxSO2. The calculations were carried out using the aug-cc-pVTZ basis set with the parameters $(\mu,\l)$ optimized on the AE6+BH6 combined set. The reference values are the non-relativistic FC CCSD(T)/cc-pVQZ-F12 values of Ref.~\onlinecite{HauKlo-JCP-12}.}
\label{tab:AE49}
\begin{tabular}{lccccccc}
\hline\hline\\[-0.1cm]
Molecule                     & RSH+CCD    & RS2H+CCD    & CCD         & RSH+RPAxSO2  & RS2H+RPAxSO2 & RPAxSO2 & Reference \\ 
%\hline                                                           
\phantom{xxxxxxxx} $(\mu,\l)=$ & (0.58,0) & (0.48,0.14) &             & (0.60,0)   & (0.48,0.34)    &        &   \\
\hline\\[-0.1cm]                                                                                                
CH                           & 79.78      & 80.67       & 81.29       & 79.70      & 81.74          & 78.30  &  83.87   \\
CH$_{2}$($^{3}$B$_{1})$      & 190.90     & 191.13      & 186.63      & 191.12     & 191.80         & 192.80 &  189.74  \\
CH$_{2}$($^{1}$A$_{1})$      & 172.28     & 173.65      & 174.30      & 172.66     & 176.76         & 188.01 &  180.62  \\
CH$_3$                       & 304.10     & 304.93      & 300.90      & 304.32     & 306.26         & 309.00 &  306.59  \\
CH$_4$                       & 412.30     & 413.61      & 409.79      & 412.71     & 415.95         & 421.50 &  418.87  \\
NH                           & 82.03      & 82.78       & 80.45       & 81.65      & 82.08          & 79.39  &  82.79   \\
NH$_2$                       & 178.75     & 179.91      & 175.44      & 178.37     & 179.41         & 177.06 &  181.96  \\
NH$_3$                       & 290.78     & 292.09      & 286.21      & 290.62     & 292.27         & 291.77 &  297.07  \\
OH                           & 105.34     & 105.52      & 102.44      & 105.24     & 105.13         & 106.41 &  106.96  \\
OH$_2$                       & 227.10     & 227.41      & 222.83      & 227.11     & 227.53         & 232.03 &  232.56  \\
FH                           & 138.56     & 138.42      & 135.71      & 138.55     & 138.27         & 143.79 &  141.51  \\
SiH$_{2}$($^{1}$A$_{1})$     & 143.87     & 145.01      & 147.74      & 145.82     & 149.59         & 159.31 &  153.68  \\
SiH$_{2}$($^{3}$B$_{1})$     & 128.70     & 129.30      & 128.36      & 130.50     & 132.24         & 136.05 &  133.26  \\
SiH$_3$                      & 219.09     & 220.24      & 221.12      & 221.04     & 224.37         & 231.15 &  228.08  \\
SiH$_4$                      & 310.76     & 312.31      & 315.62      & 313.01     & 317.89         & 328.49 &  324.59  \\
PH$_2$                       & 147.48     & 148.72      & 149.45      & 146.75     & 149.02         & 149.53 &  153.97  \\
PH$_3$                       & 229.99     & 231.72      & 232.29      & 229.92     & 233.59         & 237.54 &  241.47  \\
SH$_2$                       & 174.83     & 175.84      & 174.41      & 175.97     & 178.92         & 187.77 &  183.30  \\
ClH                          & 102.54     & 103.00      & 101.58      & 103.53     & 105.51         & 114.36 &  107.20  \\
HCCH                         & 401.05     & 402.22      & 382.78      & 401.30     & 402.39         & 400.90 &  402.76  \\
H$_2$CCH$_2$                 & 556.75     & 558.58      & 542.25      & 556.98     & 560.12         & 562.80 &  561.34  \\
H$_3$CCH$_3$                 & 703.67     & 705.79      & 692.09      & 704.35     & 708.77         & 714.48 &  710.20  \\
CN                           & 174.85     & 176.40      & 160.48      & 173.27     & 172.23         & 152.74 &  180.06  \\
HCN                          & 307.09     & 308.97      & 289.41      & 306.50     & 307.15         & 300.98 &  311.52  \\
CO                           & 255.01     & 256.25      & 238.94      & 254.90     & 255.20         & 256.53 &  258.88  \\
HCO                          & 278.43     & 279.62      & 260.11      & 277.67     & 276.99         & 272.08 &  278.28  \\
H$_2$CO                      & 369.16     & 370.55      & 353.23      & 368.87     & 369.76         & 371.37 &  373.21  \\
H$_3$COH                     & 506.70     & 507.92      & 493.19      & 506.79     & 508.43         & 512.99 &  511.83  \\
N$_2$                        & 220.09     & 222.76      & 203.11      & 218.65     & 219.07         & 208.60 &  227.44  \\
H$_2$NNH$_2$                 & 431.34     & 433.64      & 414.62      & 430.57     & 431.71         & 424.61 &  436.70  \\
NO                           & 152.57     & 154.19      & 133.80      & 151.04     & 149.36         & 137.16 &  152.19  \\
O$_2$                        & 122.94     & 124.32      & 111.15      & 120.20     & 116.90         & 92.81  &  120.54  \\
HOOH                         & 262.12     & 263.47      & 247.70      & 261.09     & 260.86         & 264.23 &  268.65  \\
F$_2$                        & 34.49      & 36.05       & 24.73       & 32.86      & 32.44          & 38.63  &  38.75   \\
CO$_2$                       & 390.04     & 390.94      & 355.41      & 389.81     & 387.23         & 381.80 &  388.59  \\
Si$_2$                       & 68.04      & 68.79       & 60.44       & 72.75      & 68.41          & 59.39  &  73.41   \\
P$_2$                        & 104.97     & 106.56      & 94.04       & 105.28     & 106.05         & 99.69  &  115.95  \\
S$_2$                        & 101.76     & 102.32      & 89.85       & 100.03     & 100.02         & 97.56  &  103.11  \\
Cl$_2$                       & 54.25      & 55.11       & 45.10       & 55.57      & 57.68          & 69.20  &  59.07   \\
SiO                          & 184.09     & 184.76      & 169.48      & 186.36     & 186.11         & 186.87 &  192.36  \\
SC                           & 162.92     & 164.51      & 148.84      & 164.34     & 166.47         & 171.52 &  170.98  \\
SO                           & 123.59     & 124.59      & 112.69      & 121.82     & 120.33         & 111.26 &  125.80  \\
ClO                          & 62.29      & 63.49       & 46.65       & 61.35      & 60.91          & 60.19  &  64.53   \\
ClF                          & 59.05      & 59.86       & 48.67       & 59.04      & 59.36          & 67.88  &  62.57   \\
Si$_2$H$_6$                  & 514.91     & 517.46      & 517.07      & 519.47     & 526.90         & 541.49 &  535.47  \\
CH$_3$Cl                     & 389.22     & 390.60      & 379.20      & 390.41     & 393.70         & 402.68 &  394.52  \\
CH$_3$SH                     & 464.50     & 466.40      & 454.91      & 465.92     & 470.11         & 479.01 &  473.49  \\
HOCl                         & 159.72     & 160.83      & 148.10      & 159.82     & 160.83         & 168.50 &  165.79  \\
SO$_2$                       & 239.81     & 242.22      & 209.59      & 239.96     & 239.55         & 237.96 &  259.77  \\ 
\hline\\[-0.1cm]                                                                                                
MAE                          & 5.76       & 4.74        & 14.52       & 5.42       & 4.22           & 6.85   &   \\
ME                           & -5.54      & -4.29       & -14.52      & -5.31      & -4.13          & -3.18  &   \\
RMSD                         & 7.15       & 6.03        & 16.93       & 6.61       & 5.33           & 9.41   &   \\
Min error                    & -20.56     & -18.01      & -50.18      & -19.81     & -20.22         & -27.73 &   \\
Max error                    &   2.40     &   3.78      & -2.34       &   1.38     &   2.06         &  10.13 &   \\
\hline\hline
\end{tabular}
\end{table*}
\endgroup
%%%%%%%%%%%%%%%%%%%%%%%%%%%%%%%%%%%%%%%%%%%%%%%%%%%%%%%%%%%%%%%%%%%%%%%%%%%%%%%%%%%%%%%%%%%%%%%%%%%%%%%%%%%%%%%%

%%%%%%%%%%%%%%%%%%%%%%%%%%%%%%%%%%%%%%%%%%%%%%%%%%%%%%%%%%%%%%%%%%%%%%%%%%%%%%%%%%%%%%%%%%%%%%%%%%%%%%%%%%%%%%%%
\begingroup
\squeezetable
\begin{table*}
\renewcommand{\arraystretch}{0.8}
\setlength{\tabcolsep}{0.07cm}
\footnotesize
\caption{Forward (F) and reverse (R) reaction barrier heights (in kcal/mol) of the DBH24/08 set calculated by RSH+CCD, RS2H+CCD, CCD, RSH+RPAxSO2, RS2H+RPAxSO2, and RPAxSO2. The calculations were carried out using the aug-cc-pVTZ basis set with the parameters $(\mu,\l)$ optimized on the AE6+BH6 combined set. The reference values are taken from Ref.~\onlinecite{ZheZhaTru-JCTC-09}.}
\label{tab:DBH24}
\begin{tabular}{lccccccc}
\hline\hline\\[-0.1cm]
Reaction                                      & RSH+CCD     & RS2H+CCD    & CCD         & RSH+RPAxSO2 & RS2H+RPAxSO2 & RPAxSO2 & Reference   \\
\phantom{xxxxxxxxxxxxxxxxx} $(\mu$,$\l)=$     & (0.58,0)    & (0.48,0.14) &             & (0.60,0)    & (0.48,0.34)  &         & \\
\hline\\[-0.1cm]
                                              &   F/R       &  F/R        &   F/R       & F/R         & F/R          & F/R     & \\
\\
Heavy-atom transfer                           &             &             &             &              \\
H + N$_{2}$O $\rightarrow$ OH + N$_{2}$       & 15.37/74.77 & 14.82/73.49 & 21.95/95.64 & 19.48/78.75 & 19.93/81.70 & 32.45/103.17 & 17.13/82.47 \\
H + ClH $\rightarrow$ HCl + H                 & 16.63/16.63 & 16.10/16.10 & 23.66/23.66 & 19.63/19.63 & 19.40/19.40 & 23.94/23.94  & 18.00/18.00\\      
CH$_{3}$ + FCl $\rightarrow$ CH$_{3}$F + Cl   &  4.23/60.09 & 3.41/58.36  & 11.35/68.10 & 8.57/63.98  & 9.00/63.24  & 20.68/69.49  & 6.75/60.00 \\      
\\
Nucleophilic substitution                     &             &             &             &             &             &          &  \\
Cl$^{-}$$\cdots$CH$_{3}$Cl $\rightarrow$ ClCH$_{3}$$\cdots$Cl$^-$ & 15.08/15.08  & 14.33/14.33 & 15.90/15.90 & 14.82/14.82   & 14.20/14.20    & 14.65/14.65 & 13.41/13.41 \\  
F$^{-}$$\cdots$CH$_{3}$Cl $\rightarrow$ FCH$_{3}$$\cdots$Cl$^{-}$ & 4.30/31.30   & 3.72/30.50  & 5.00/33.55  & 4.22/31.28    & 3.72/31.14     & 4.19/32.83  &  3.44/29.42  \\
OH$^{-}$ + CH$_{3}$F $\rightarrow$ HOCH$_{3}$ + F$^{-}$           & -2.04/20.70  & -2.81/19.27 & 1.32/21.59  & -1.93/20.70   & -2.00/19.63    & 0.37/20.59  & -2.44/17.66 \\
\\
Unimolecular and association                      &              &             &             &              &             &             &       \\
H + N$_{2}$ $\rightarrow$ HN$_{2}$                & 11.39/11.33  & 10.98/11.00 & 19.05/12.02 & 13.78/8.69   & 14.18/12.78 & 23.28/12.49 & 14.36/10.61  \\
H + C$_{2}$H$_{4}$ $\rightarrow$ CH$_{3}$CH$_{2}$ & -0.08/42.59  & 0.07/42.57  & 4.56/45.59  & 2.60/45.39   & 2.85/45.56  & 7.67/49.89  & 1.72/41.75  \\
HCN $\rightarrow$ HNC                             & 48.47/34.39  & 47.62/33.50 & 49.24/34.87 & 48.34/34.58  & 47.51/33.79 & 46.87/34.48 & 48.07/32.82 \\
\\
Hydrogen transfer                                &                &             &             &             &             &             &    \\
OH + CH$_{4}$ $\rightarrow$ CH$_{3}$ + H$_{2}$O  & 4.11/17.72     & 3.21/16.51  & 12.20/24.08 & 5.91/19.50  & 6.45/19.25  & 11.60/24.82 & 6.70/19.60  \\
H + OH $\rightarrow$ O + H$_{2}$                 & 11.16/8.19     & 9.97/7.17   & 14.91/19.92 & 12.83/9.83  & 11.94/11.25 & 15.56/19.03 & 10.70/13.10 \\
H + H$_{2}$S $\rightarrow$ H$_{2}$+ HS           & 3.02/14.39     & 2.61/14.13  & 6.20/20.36  & 4.23/14.75  & 4.02/15.11  & 6.24/17.55  & 3.60/17.30  \\
\hline\\[-0.1cm]                                 
MAE                                              &  1.91          & 2.08        & 4.29  & 1.78    & 1.37     & 5.63   &       \\
ME                                               &  -0.78         & -1.53       & 4.29  & 0.70    & 0.86     & 5.53   &       \\
RMSD                                             &  2.51          & 2.86        & 4.97  & 2.09    & 1.68     & 7.52   &        \\
Min error                                        &  -7.70         & -8.98       & 1.17  & -3.72   & -2.19    & -1.20  &        \\
Max error                                        &  3.04          & 1.61        & 13.17 & 3.98    & 3.81     & 20.70  &       \\ 
\hline\hline
\end{tabular}
\end{table*}
\endgroup
%%%%%%%%%%%%%%%%%%%%%%%%%%%%%%%%%%%%%%%%%%%%%%%%%%%%%%%%%%%%%%%%%%%%%%%%%%%%%%%%%%%%%%%%%%%%%%%%%%%%%%%%%%%%%%%%

%%%%%%%%%%%%%%%%%%%%%%%%%%%%%%%%%%%%%%%%%%%%%%%%%%%%%%%%%%%%%%%%%%%%%%%%%%%%%%%%%%%%%%%%%%%%%%%%%%%%%%%%%%%%%%%%
\begingroup
\squeezetable
\begin{table*}
\renewcommand{\arraystretch}{0.8}
\setlength{\tabcolsep}{0.03cm}
\footnotesize
\caption{Interaction energies (in kcal/mol) for the complexes of the A24 set calculated by RSH+CCD, RS2H+CCD, CCD, RSH+RPAxSO2, RS2H+RPAxSO2, and RPAxSO2. The calculations were carried out using the aug-cc-pVTZ basis set with the counterpoise correction and using the parameters $(\mu,\l)$ optimized on the AE6+BH6 combined set. The reference values are the non-relativistic reference interaction energies from Ref.~\onlinecite{RezDubJurHob-PCCP-15}.
}
\label{tab:A24}
\begin{tabular}{lcccccccccc}
\hline\hline\\[-0.1cm]
Complex                          & RSH+MP2& RS2H+MP2 & MP2 & RSH+CCD & RS2H+CCD  & CCD      & RSH+RPAxSO2 & RS2H+RPAxSO2 & RPAxSO2 & Reference     \\
\phantom{xxxxxxx}      $(\mu$,$\l)=$  & (0.58,0) & (0.46,0.58) &    & (0.58,0) & (0.48,0.14) &             & (0.60,0)    & (0.48,0.34)  &         & \\
\hline \\[-0.1cm]  
\multicolumn{10}{l}{Hydrogen bonds}\\[0.1cm]
water...ammonia $\Cs$            & -7.049 & -6.834 &   -6.303 & -7.059 & -7.113 & -5.853  &  -7.020 & -6.927  & -6.012 & -6.555 \\
water dimer     $\Cs$            & -5.443 & -5.207 &   -4.727 & -5.465 & -5.473 & -4.492  &  -5.435 & -5.323  & -4.622 & -5.049 \\
HCN dimer       $\Cs$            & -5.305 & -5.109 &   -4.783 & -5.131 & -5.109 & -4.479  &  -5.166 & -5.058  & -4.644 & -4.776 \\
HF dimer        $\Cs$            & -4.968 & -4.712 &   -4.194 & -5.004 & -5.001 & -4.126  &  -4.978 & -4.870  & -4.247 & -4.601 \\
ammonia dimer   $\Cdh$           & -3.230 & -3.186 &   -3.007 & -3.268 & -3.310 & -2.732  &  -3.250 & -3.218  & -2.828 & -3.170 \\
\hline\\[-0.1cm]
MAE                              & 0.369  & 0.179   & 0.230  &  0.355  & 0.371   & 0.494  &  0.340  & 0.249   & 0.360  &        \\[0.1cm]
\hline\\[-0.1cm]

\multicolumn{10}{l}{Mixed electrostatics/dispersion}\\[0.1cm]
HF...methane   $\Ctv$            & -1.823 & -1.711  &  -1.494 & -1.880 & -1.895 & -1.298  &  -1.871 & -1.803  &-1.395  & -1.664 \\
ammonia...methane $\Ctv$         & -0.772 & -0.746  &  -0.663 & -0.812 & -0.820 & -0.582  &  -0.809 & -0.785  &-0.648  & -0.779 \\
water...methane   $\Cs$          & -0.678 & -0.648  &  -0.579 & -0.717 & -0.717 & -0.511  &  -0.711 & -0.680  &-0.559  & -0.681 \\
formaldehyde dimer $\Cs$         & -5.525 & -4.990  &  -4.205 & -5.291 & -5.239 & -3.565  &  -5.357 & -5.058  &-4.053  & -4.524 \\
water...ethene $\Cs$             & -2.837 & -2.771  &  -2.608 & -2.796 & -2.829 & -2.181  &  -2.810 & -2.759  &-2.373  & -2.586 \\
formaldehyde...ethene $\Cs$      & -1.831 & -1.733  &  -1.578 & -1.763 & -1.765 & -1.255  &  -1.788 & -1.707  &-1.443  & -1.634 \\ 
ethyne dimer  $\Cdv$             & -1.694 & -1.676  &  -1.570 & -1.600 & -1.620 & -1.307  &  -1.629 & -1.616  &-1.484  & -1.535 \\ 
ammonia...ethene $\Cs$           & -1.472 & -1.465  &  -1.427 & -1.448 & -1.470 & -1.121  &  -1.459 & -1.430  &-1.246  & -1.395 \\
ethene dimer $\Cdv$              & -1.187 & -1.182  &  -1.191 & -1.139 & -1.161 & -0.762  &  -1.158 & -1.105  &-0.920  & -1.109 \\
methane...ethene $\Cs$           & -0.518 & -0.515  &  -0.515 & -0.524 & -0.532 & -0.375  &  -0.529 & -0.508  &-0.435  & -0.518 \\
\hline\\[-0.1cm]
MAE                              & 0.193  & 0.115   & 0.094  &  0.155  & 0.162   & 0.347  &  0.170  & 0.106   & 0.187  &        \\[0.1cm]
\hline\\[-0.1cm]

\multicolumn{10}{l}{Dispersion dominated}\\[0.1cm]
borane...methane $\Cs$           &  -1.472&   -1.422&  -1.304 & -1.627 & -1.681 & -0.966  &  -1.601 & -1.527  &-1.048  & -1.521 \\
methane...ethane $\Cs$           &  -0.786&   -0.766&  -0.746 & -0.827 & -0.835 & -0.554  &  -0.823 & -0.776  &-0.633  & -0.844 \\
methane...ethane $\Cs$           &  -0.546&   -0.531&  -0.511 & -0.597 & -0.602 & -0.401  &  -0.594 & -0.559  &-0.460  & -0.617 \\
methane dimer $\Dtd$             &  -0.475&   -0.465&  -0.455 & -0.523 & -0.528 & -0.354  &  -0.518 & -0.487  &-0.399  & -0.542 \\
Ar...methane  $\Ctv$             &  -0.386&   -0.375&  -0.359 & -0.395 & -0.403 & -0.241  &  -0.390 & -0.365  &-0.272  & -0.403 \\
Ar...ethene   $\Cdv$             &  -0.373&   -0.370&  -0.374 & -0.348 & -0.362 & -0.214  &  -0.345 & -0.327  &-0.243  & -0.354 \\ 
ethene...ethyne $\Cdv$           &  0.761 &   0.719 &  0.590  & 0.907  & 0.861  & 1.203   &  0.919  & 0.960   &1.189   & 0.801  \\
ethene dimer $\Ddh$              &  0.939 &   0.917 &  0.796  & 1.042  & 0.990  & 1.361   &  1.068  & 1.109   &1.350   & 0.909  \\
ethyne dimer $\Ddh$              &  1.029 &   0.963 &  0.808  & 1.213  & 1.172  & 1.482   &  1.216  & 1.263   &1.494   & 1.096  \\
\hline\\[-0.1cm]
MAE                              & 0.046  & 0.067   & 0.132  &  0.059  & 0.047   & 0.310  &  0.063  & 0.086   & 0.273  &        \\[0.1cm]
\hline\\[-0.1cm]

Total MAE                        & 0.175  & 0.111   & 0.136  &  0.161  & 0.163   & 0.364  &  0.165  & 0.128   & 0.255  &        \\ 
Total ME                         & -0.150 & -0.074  & 0.069  &  -0.125 & -0.141  & 0.364  &  -0.124 & -0.063  & 0.255  &        \\ 
Total RMSD                       & 0.289  & 0.155   & 0.176  &  0.249  & 0.249   & 0.410  &  0.255  & 0.185   & 0.293  &        \\ 
Total MA\%E                      & 7.2\%  & 6.4\%   & 9.6\%  & 6.7\%  & 6.5\%    & 26.4\% & 7.2\%   & 7.0\%   & 19.6\%  &        \\
\hline
\hline
\end{tabular}
\end{table*}
\endgroup
%%%%%%%%%%%%%%%%%%%%%%%%%%%%%%%%%%%%%%%%%%%%%%%%%%%%%%%%%%%%%%%%%%%%%%%%%%%%%%%%%%%%%%%%%%%%%%%%%%%%%%%%%%%%%%%%

\section{Results and discussion}
\label{sec:results}

\subsection{Optimization of the parameters on the AE6 and BH6 sets}

We start by applying the RS2H+CCD and RS2H+RPAxSO2 methods on the small AE6 and BH6 sets for determining optimal values for the parameters $\mu$ and $\l$. Figure~\ref{AE6_BH6} shows the MAEs for these two sets as a function of $\l$ for $\mu=0.5$, which is close to the optimal value of $\mu$ for range-separated hybrids~\cite{GerAng-CPL-05a,MusReiAngTou-JCP-15}. Note that, particularly for the AE6 set, the MAEs near the $\l=1$ end of the curves, corresponding to full-range CCD and RPAxSO2, may not be well converged with the cc-pVTZ basis set since these full-range methods have a slow convergence with the size of the basis set~\cite{MusReiAngTou-JCP-15}. Since we will be interested in practice in the RS2H+CCD and RS2H+RPAxSO2 methods with relatively small values of $\lambda$ for which basis convergence is expected to be fast~\cite{KalTou-JCP-18}, we did not think necessary to use larger basis sets in this study.

For the AE6 set, with the RS2H+CCD method, a minimal MAE of about 4 kcal/mol is obtained close to the $\l=0$ end of the curve (corresponding to RSH+CCD). For $\l \gtrsim 0.2$, the MAE obtained with RS2H+CCD increases rapidly with $\l$, reaching a maximal MAE of about 22 kcal/mol for $\l=1$ (corresponding to full-range CCD). In comparison with RS2H+CCD, the RS2H+RPAxSO2 method always gives smaller MAEs. A minimal MAE of about 3.5 kcal/mol is obtained in a remarkably large range of $\l$ between about 0.1 and 0.8. For the BH6 set, the two MAE curves display marked minima at an intermediate value of $\l$. The RS2H+CCD method gives a minimal MAE of about 2 kcal/mol for $\l\approx 0.75$, while the RS2H+RPAxSO2 method gives a minimal MAE of about 1 kcal/mol for $\l\approx 0.4$. Clearly, in this RSDH scheme, simplifying the CCD ansatz by making the ring approximation (with exchange terms) is actually largely beneficial.

We have also determined optimal values of $\mu$ and $\l$ that minimize the total MAE of the combined AE6 + BH6 set, and which could be used for general chemical applications. For the RS2H+CCD method, the optimal parameter values are $(\mu,\l)=(0.48,0.14)$. For the RS2H+RPAxSO2 method, the optimal parameter values are $(\mu,\l)=(0.48,0.34)$. Note that for the RS2H+MP2 method, the optimal parameter values determined in Ref.~\onlinecite{KalTou-JCP-18} were $(\mu,\l)=(0.46,0.58)$. Thus, the value of $\l$ is more sensitive to the wave-function method used than the value of $\mu$. The decrease of the optimal value of $\l$ in the series RS2H+MP2 $\to$ RS2H+RPAxSO2 $\to$ RS2H+CCD is consistent with the deterioration of the accuracy of the atomization energies in the corresponding full-range series MP2 $\to$ RPAx-SO2 $\to$ CCD. As mentioned above, the fact that the optimal values of $\l$ obtained with RS2H+CCD and RS2H+RPAxSO2 are small is advantageous for basis convergence since only a small fraction of the short-range interaction is treated by the correlated wave-function method~\cite{KalTou-JCP-18}. In the following, we further assess the methods with the determined optimal parameters.

\subsection{Assessment on the AE49 and DBH24/08 sets of atomization energies and reaction barrier heights}

We assess now the RS2H+CCD and RS2H+RPAxSO2 methods, evaluated with the previously determined optimal parameters $(\mu,\l$), on the larger AE49 and DBH24/08 sets of atomization energies and reaction barrier heights. The results are reported in Tables~\ref{tab:AE49} and~\ref{tab:DBH24}, and compared with other methods corresponding to limit cases of these RSDH schemes: RSH+CCD and RSH+RPAxSO2 (corresponding to the $\l=0$ limit) and full-range CCD and RPAxSO2 (corresponding to the $\l=1$ limit). Again, for atomization energies, one should bear in mind that the full-range CCD and RPAxSO2 results may not be well converged with the aug-cc-pVTZ basis set~\cite{MusReiAngTou-JCP-15}, even though this basis set seems sufficient to reveal the inaccuracy of the atomization energies obtained with these full-range methods.

On the AE49 set, RS2H+CCD gives a MAE of 4.7 kcal/mol, which is a small improvement over RSH+CCD (MAE of 5.8 kcal/mol) but a large improvement over full-range CCD (MAE of 14.5 kcal/mol). Similarly, RS2H+RPAxSO2 with a MAE of 4.2 kcal/mol provides an improvement over both RSH+RPAxSO2 (MAE of 5.4 kcal/mol) and full-range RPAxSO2 (MAE of 6.9 kcal/mol). Comparing with the results obtained with RS2H+MP2 method in Ref.~\onlinecite{KalTou-JCP-18}, we see that RS2H+RPAxSO2 provides overall a similar accuracy for atomization energies.

On the DBH24/08 set, RS2H+CCD gives a MAE of 2.1 kcal/mol, similar to RSH+CCD (MAE of 1.9 kcal/mol) but in large improvement over full-range CCD (MAE of 4.3 kcal/mol). Again, RS2H+RPAxSO2 gives the smallest MAE of 1.4 kcal/mol, comparable to RSH+RPAxSO2 (MAE of 1.8 kcal/mol) but in large improvement over full-range RPAxSO2 (MAE of 5.6 kcal/mol). Also here, in comparison with the results obtained with the RS2H+MP2 method in Ref.~\onlinecite{KalTou-JCP-18}, we see that RS2H+RPAxSO2 provides a roughly similar accuracy for reaction barrier heights.

\subsection{Assessment on the A24 set of intermolecular interactions and on the benzene dimer}

We test now the RS2H+CCD and RS2H+RPAxSO2 methods on weak intermolecular interactions. Table~\ref{tab:A24} reports the interaction energies for the 24 complexes of the A24 set calculated by RSH+CCD, RS2H+CCD, CCD, RSH+RPAxSO2, RS2H+RPAxSO2, and RPAxSO2. For comparison, we also report results obtained with the RSH+MP2, RS2H+MP2, and MP2 methods since this A24 set was not considered in Ref.~\onlinecite{KalTou-JCP-18}.

Let us start by discussing the results obtained with the full-range methods. Full-range CCD and RPAxSO2 largely underbind all the complexes, with MA\%Es of about 26 \% and 20 \%, respectively. Part of this underestimation of the interaction energies could be due to the basis-set incompleteness. Nevertheless, our results turn out to be quite similar to the results obtained with other RPAx variants, namely eh-TDHF and AC-SOSEX, tested in Ref.~\onlinecite{DixClaLebRoc-JCTC-17} using PBE orbitals with a plane-wave basis set. Other authors have found that other RPAx variants based on the local exact-exchange kernel perform significantly better~\cite{HelColGir-PRB-18}.

We discuss now the results obtained with the range-separated methods. As found in Ref.~\onlinecite{KalTou-JCP-18}, in comparison with RSH+MP2, the RS2H+MP2 method provides a substantial and systematic improvement for hydrogen-bond complexes, and a smaller overall improvement for complexes with mixed electrostatic/dispersion interactions. However, for the small dispersion-dominated complexes considered in this set, RS2H+MP2 appears to be slightly less accurate than RSH+MP2. The RS2H+CCD method does not provide any overall improvement over RSH+CCD. This may be due to the very small value of $\l$ used in RS2H+CCD, which in addition is compensated by a smaller value of $\mu$ in comparison to RSH+CCD. Similarly to RS2H+MP2, the RS2H+RPAxSO2 method provides a small and systematic improvement over RSH+RPAxSO2 for hydrogen-bond and mixed complexes, but a small deterioration over RSH+RPAxSO2 for the small dispersion-dominated complexes considered in this set. Overall, RSH+MP2, RS2H+MP2, RSH+CCD, RS2H+CCD, RSH+RPAxSO2, and RS2H+RPAxSO2 all give total MA\%Es of around 7\% and it is thus hard to discriminate between them based on this A24 set.

\begin{figure}[t]
\centering
\includegraphics[scale=0.32,angle=-90]{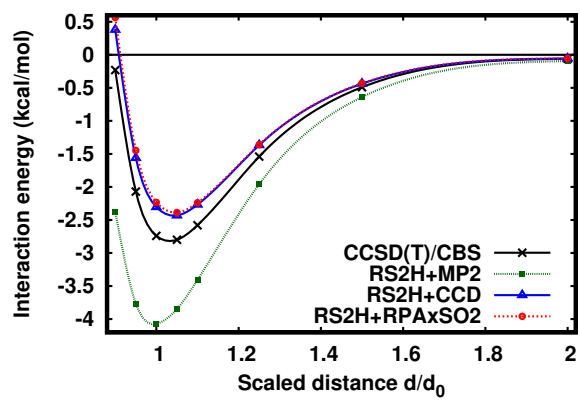}
\caption{Interaction energy curve of the benzene dimer in the stacked configuration calculated by RS2H+MP2, RS2H+CCD, and RS2H+RPAxSO2 as a function of the scaled distance between the monomers $d/d_0$ where $d_0$ is a fixed distance. The calculations were carried out using the aug-cc-pVDZ basis set with the counterpoise correction and using the parameters $(\mu,\l)$ optimized on the AE6+BH6 combined set. The geometries and the CCSD(T)/CBS reference interaction energies are from the S66$\times$8 data set (item number 24) of Ref.~\onlinecite{RezRilHob-JCTC-11}.}
\label{benzene_dimer}
\end{figure}

Finally, we consider in Figure~\ref{benzene_dimer} the interaction energy curve of the benzene dimer in the stacked (parallel-displaced) configuration, the simplest prototype of aromatic $\pi$-$\pi$ intermolecular interactions. For this system, it is known that, in the CBS limit, MP2 considerably overbinds and going to CCSD(T) is necessary to obtain accurate interaction energies~\cite{SinValShe-JACS-02,SinShe-JPCA-04}. Similarly to full-range MP2, we find that the RS2H+MP2 method largely overbinds, with an equilibrium interaction energy too low by about 1.3 kcal/mol. The RS2H+CCD and RS2H+RPAxSO2 methods, which give very similar interaction energy curves, moderately underbind, with an equilibrium interaction energy underestimated by about 0.4 kcal/mol. We thus see that, for this type of system, it is advantageous to supplant MP2 by CCD or RPAxSO2 in the RSDH scheme.

%%%%%%%%%%%%%%%%%%%%%%%%%%%%%%%%%%%%%%%%%%%%%%%%%%%%%%%%%%%%%%%%%%%%%%%%%%%%%%%%%%%%%%%%%%%%%%%%%%
%%%%%%%%%%%%%%%%%%%%%%%%%%%%%%%%%%%%%%%%%%%%%%%%%%%%%%%%%%%%%%%%%%%%%%%%%%%%%%%%%%%%%%%%%%%%%%%%%%

\section{Conclusion}
\label{sec:conclusion}

We have constructed CCD/DFT and RPA/DFT hybrid approximations using the RSDH scheme which is based on a two-parameter CAM-like decomposition of the electron-electron interaction. In comparison with the previously existing RSH+CCD and RSH+RPAxSO2 range-separated hybrids, the present RS2H+CCD and RS2H+RPAxSO2 methods incorporates a fraction $\l$ of short-range electron-electron interaction in the wave-function part. Tests on atomization energies, reaction barrier heights, and weak intermolecular interactions show that this addition of short-range interaction is globally beneficial for RS2H+RPAxSO2, while the effect is less important for RS2H+CCD. In comparison with the simpler RS2H+MP2 method, the RS2H+RPAxSO2 method is globally as accurate for atomization energies, reaction barrier heights, and weak intermolecular interactions of small molecules. For the more complicated case of the benzene dimer in the stacked configuration, RS2H+RPAxSO2 reduces the large overbinding obtained with the RS2H+MP2 method. Even though more tests should now be performed on larger systems, if we had to recommend a computational method for general chemical applications among the methods tested in this work, it would thus be RS2H+RPAxSO2 with parameters $(\mu,\l)=(0.48,0.34)$. More generally, we hope that the formalism provided in the present work will be useful for constructing more beyond-MP2 double-hybrid approximations with minimal empiricism.

\section*{Acknowledgements}
We thank Labex MiChem for having provided PhD financial support for C. Kalai.

% BIBLIOGRAPHY---------------------------------------------
%\bibliographystyle{jchemphys}
%\bibliography{BibTeX1,biblio}

\end{document}